# Functional Magnetic Resonance Spectroscopy in the mouse

Clémence Ligneul, Francisca F. Fernandes and Noam Shemesh[*]

*Champalimaud Research, Champalimaud Centre for the Unknown, Lisbon, Portugal*

**\*Corresponding author:**
Dr. Noam Shemesh, Champalimaud Research, Champalimaud Centre for the Unknown, Av. Brasilia 1400-038, Lisbon, Portugal**.** E-mail: noam.shemesh@neuro.fchampalimaud.org . Phone number: +351 210 480 000 ext. #4467.

**Running title:** Metabolic changes upon visual stimulation in the mouse detected using fMRS

# Abstract

Functional magnetic resonance spectroscopy (fMRS) quantifies metabolic variations upon presentation of a stimulus and can therefore provide complementary information compared to functional magnetic resonance imaging (fMRI). However, to our knowledge, fMRS has not yet been performed in the mouse, despite that murine models are crucial for basic and applied research. Here, we performed fMRS experiments in the mouse, for the first time, and show the feasibility of such an approach for reliably quantifying metabolic variations. In particular, we observed metabolic variations in the superior colliculus of mice upon visual stimulation in a block paradigm commonly used for fMRI (short periods of stimulus), followed by a recovery period. We notably report a robust modulation of glutamate, as well as a modulation of NAAG, PCr and Cr. A control experiment with no stimulation reveals potential metabolic signal "drifts" that are not correlated with the functional activity, which should be taken into account when analyzing fMRS data in general. Our findings are promising for future applications of fMRS in the mouse.

# Highlights

- Functional magnetic resonance spectroscopy is feasible in the sedated mouse.
- A modulation of Glutamate is robustly observed in the mouse superior colliculus upon visual stimulation.
- A GLM regression indicates that glutamate seems to be upregulated during the block-paradigm but does not exhibit stimulus-locked variation with this design.

# Introduction

The complexity of brain function calls for diverse and complementary approaches capable of reporting on different aspects of neural activity. Functional magnetic resonance imaging (fMRI) is one of the most popular noninvasive techniques capable of studying human brain activity from a global brain perspective and with relatively high spatial resolution. Relying on neurovascular couplings as surrogate reporters for activity (Logothetis *et al.*, 2001), fMRI can identify brain areas involved in a given task or in processing information. At higher fields, Blood Oxygen Level Dependent (BOLD) fMRI even allows for laminar resolution (Polimeni *et al.*, 2010; Goense, Merkle and Logothetis, 2012), and, to some extent, specificity and insight into how the brain processes stimuli and integrates information.

Despite its invaluable contributions to cognitive neuroscience and to clinical applications, BOLD fMRI is typically less specific in characterizing neural activity, mainly because of its indirect nature (Logothetis, 2008). For example, neither excitation and inhibition balances nor neuromodulatory effects can be specifically inferred from (typical) BOLD responses.

Magnetic resonance spectroscopy (MRS) can provide noninvasive information on myriad molecules involved in brain metabolism if they are sufficiently concentrated (~mM) *in vivo*, thus endowing MRS with some level of specificity. Certain metabolites are concentrated in particular cell types (Urenjak *et al.*, 1993; Belle *et al.*, 2002; Griffin *et al.*, 2002), making MRS more specific towards intracellular space compared to its water-driven MRI counterpart. Particularly interesting perhaps are MRS's capabilities of probing glutamate (Glu) and γ-aminobutyric acid (GABA) – the main excitatory and inhibitory neurotransmitters in the brain, respectively. In addition, molecules involved in prominent metabolic pathways, such as glucose (Glc) and lactate (Lac), can be accessed using high-fidelity MRS. However, given the ~$10^4$ lower concentration of metabolites compared to water, MRS has a much lower sensitivity compared to water-based MRI, usually leading to long acquisition times and poor spatial resolution. Increasingly higher magnetic fields provide increasingly higher signal-to-noise, thereby overcoming some of the sensitivity limitations; furthermore, higher magnetic fields also allow better spectral resolution, which can in turn increase the reliability of metabolite quantification (Mlynárik *et al.*, 2008). However, the increased spectral bandwidths associated with higher fields also require high-power RF pulses or other alternatives (e.g. adiabatic pulses), which can then contribute to suboptimal SNR or increased SAR.

Given the abovementioned challenges with MRS, it is perhaps not surprising that functional MRS (fMRS) experiments – i.e., MRS experiments designed to probe metabolites with high temporal resolution such that their variations during a task or upon a stimulus could be measured – remain quite rare. In humans, Mangia et al. (Mangia, Tkáč, Gruetter, *et al.*, 2007; Mangia, Tkáč, Logothetis, *et al.*, 2007) and then others (Lin *et al.*, 2012; Schaller *et al.*, 2014; Bednařík *et al.*, 2015, 2018; Taylor *et al.*, 2015; Betina Ip *et al.*, 2017; Cao *et al.*, 2017; Mekle *et al.*, 2017; Martínez-Maestro, Labadie and Möller, 2018; Boillat *et al.*, 2019) harnessed the increasing availability of high field 7 T scanners to record metabolic variations upon visual stimulation, often interpreted as a result of increased oxidative metabolism upon neural activity. Koush et al. (Koush *et al.*, 2019) were able to probe lactate variations using J-Edited fMRS at 4 T. At more conventional fields of 3 T, fMRS acquisitions were carried out following some task-based stimuli (Apšvalka *et al.*, 2015; Kühn *et al.*, 2016), but also using acute pain stimuli (Gussew *et al.*, 2010; Cleve, Gussew and Reichenbach, 2015) for which associated metabolic variations are larger (Mullins, 2018). In a majority of studies, increases in Glu (1-5%) (Mullins, 2018) and lactate were evidenced in the active period. When spectral quantification was reliable enough, decreases in glucose and aspartate (Asp) levels were reported (Lin *et al.*, 2012; Bednařík *et al.*, 2015), as well as stimulus-dependent modulations in GABA levels (Cleve, Gussew and Reichenbach, 2015).

Until now, the typical fMRS paradigms involved sustained (> 5 min) sensory stimulus, to facilitate signal averaging for robust quantification of metabolites during the active/resting states and to reach a metabolic steady state. However, habituation may play an important role in such sustained stimuli. In addition, the dynamics of such slow fMRS experiments can be difficult to interpret. Preclinical experiments, with potentially much higher magnetic fields and their associated higher spectral resolution, as well as the availability of transgenic lines and possibility of performing more invasive validation experiments, can be extremely valuable to promote a better understanding and interpretation of fMRS. The first preclinical $^1$H-fMRS experiments were conducted in the rat (Just *et al.*, 2013) still using sustained (> 10 min) stimuli, where increases in Glu and Lac and decreases in Glc could be observed despite the sedation. Since then, only very few other studies (Just and Sonnay, 2017; Sonnay, Duarte and Just, 2017; Sonnay *et al.*, 2018; Just and Faber, 2019) have been performed in the rat and the tree shrew, generally reproducing the initial findings.

The ease of manipulating the mouse genome and the consequent vast array of transgenic lines make mouse models highly important for basic research at large and for MRS research in particular. Optogenetic and chemogenetic mouse lines probing different cell types and circuits are available, as are many animal models of disease. However, insofar, fMRS studies have not been conducted in the mouse, to our knowledge. This can probably be traced to difficulties associated with the very small voxel sizes required for achieving regional specificity in mouse MRS (Tkáč *et al.*, 2004), which would significantly reduce sensitivity. The difficulty is exacerbated by the need of a stable sedation (Grandjean *et al.*, 2014), which can be more challenging in mice than in rats.

In this study, we report the first – to our knowledge – functional $^1$H MRS experiments in the sedated mouse. As a simple proof of principle, we use visual stimulation and choose the superior colliculus (SC) as the target voxel, as it receives inputs directly from the optic tract and shows very robust BOLD responses upon simple visual stimuli. We use a relatively common preclinical field (9.4 T) but perform the experiments using a cryoprobe, which provides a facilitating ~2.5 increase in sensitivity compared with conventional reception coils. Furthermore, rather than using a sustained stimulus, we design a block paradigm with relatively short periods of stimulation (24s), that can alleviate adaptation and probe a different dynamic range which can potentially complement information arising from (water based) fMRI responses. Our observations of robust fMRS responses are a steppingstone for future studies aiming to enhance the specificity of fMRI experiments in normal and diseased mice.

## Material and methods

All animal experiments were preapproved by the institutional and national authorities and carried out according to European Directive 2010/63.

**Animal sedation and preparation**

Anesthesia in mice (N = 20 C57Bl6 females, 3-4 months old, weighting 24-26g) was induced using ~5% isoflurane. The percentage of isoflurane was progressively decreased during animal preparation (namely, placing the animals on the animal bed and positioning the mouse using a bite bar and ear bars, medetomidine subcutaneous cannula and rectal probe insertion). The anesthesia regime was then switched to a mix of low dose isoflurane (~0.3%) in a 28% $O_2$/air mixture and subcutaneous medetomidine infusion (bolus: 0.4mg/kg, constant infusion: 0.8mg/kg/h, optimized from (Grandjean *et al.*, 2014)). Temperature (maintained at 36.0±0.5°C) and respiration rate were monitored throughout the experiments and remained stable once the functional spectroscopy session began.

**Experimental design**

*Magnets and coils*. All data were acquired using a 9.4 T Bruker BioSpec scanner (Bruker, Karlsruhe, Germany) interfaced with an AVANCE III HD console driving a gradient system capable of producing up to 660 mT/m isotropically. The scanner runs on Paravision 6.0.1 software (Bruker, Ettlingen, Germany). An 86 mm volume quadrature resonator scanner (Bruker, Karlsruhe, Germany) was used for transmission and a 4-elements array cryoprobe (Bruker, Fallanden, Switzerland) was used for signal reception.

*Setup for delivering visual stimuli.* Optic fibers delivering a flashing blue light generated by a LED (wavelength = 470 nm, intensity = 0.8 W/m$^2$, frequency = 4 Hz, pulse duration = 15 ms) were positioned horizontally on each side of the mouse bed at about 1cm to the eye.

*Stimulation paradigm*

The block paradigm (BP) alternated between 48 sec without stimulation (OFF) and 24 sec with visual stimulation (ON) periods. This was repeated 5 times, followed by a single OFF period. For functional MRS acquisitions, the BP was followed by a recovery (R) period of the same duration (6 min 48 sec) for a total duration of 13 min 36 sec per BP-R paradigm.

*MRS pulse sequence adaptation*

Metabolites ranging from 1 to 4 ppm were targeted and observed in this study. Therefore, to ensure that the functional signals arise from the same voxel, chemical shift displacements needed to be minimized. To this end, a LASER (Localization by Adiabatic SElective Refocusing, (Garwood and Delabarre, 2001)) sequence was implemented and adapted for fMRS acquisitions in this study. In particular, the phase cycling of the LASER sequence, usually counting 32 to 64 steps for a good spurious signal removal (Kingsley, 1994), was reduced to 4 steps (excitation {0 2 1 3}, refocusing {1 3 0 2), receiver {0 2 1 3}) to allow a better temporal resolution. To compensate for the suboptimal phase cycling, 4 blocks of outer volume suppression (OVS) module were added prior to signal excitation.

*Timeline of experiments*

**Figure 1A** describes the principal experimental timeline for the functional experiment, which was designed to ensure robust visual activation throughout its entire duration. Below, we briefly mention each step and its role in the experimental design. Detailed parameters for pulse sequences used in each phase within the timeline are presented in the following section.

    **Step 1: Blood Oxygenation Level Dependent (BOLD) fMRI (~7 min).** Following routine adjustments, and a well resolved anatomical image for slices and voxel positioning, this simple experiment

was performed to ensure robust activation in the superior colliculus at the beginning of the experiment, following the BP.

**Step 2: fMRS acquisitions (~1 h 10 min).** Once a detectable BOLD-response was ascertained in the superior colliculus with a rapid analysis in PV6.0.1, we proceeded with the MRS acquisitions on upfield metabolites (with the water suppression module), following the BP-R, repeated 5 times.

**Step 3: Repeated measurement of BOLD fMRI (~7 min).** To ensure that habituation to stimulus in the superior colliculus is minimal, Step 1 was repeated and BOLD fMRI maps were measured again.

**Step 4: fMRS acquisition on the water signal (~7 min).** To get the precise timing of vascular effects in the voxel of interest and correct for the vascular effect on metabolites during post-processing, identical fMRS experiments were performed on the water signal (by turning off the water suppression module). This also verified the voxel positioning and informed about LASER-driven water BOLD responses.

**Figure 1B** describes the control experiment. The timeline is identical to the functional experiment's timeline, except that no visual stimulation is delivered during Step 2, and Step 4 was not acquired (since vascular correction is irrelevant). For comparison with the above-mentioned BP-R paradigm, the "paradigm" of Step 2 for the control experiment was divided into two periods of 6 min 48 sec, that we termed "No Stim 1" and "No Stim 2".

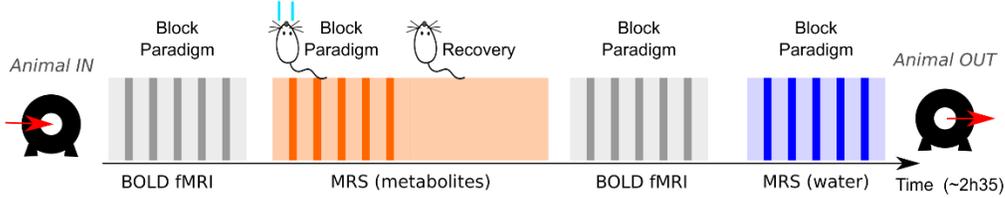

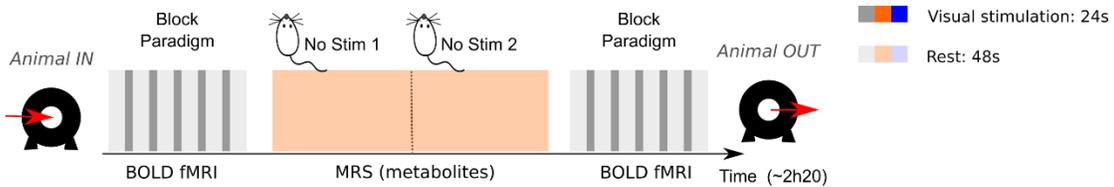

**Figure 1**. Representative timelines of the functional MRS experiment **(A)** and the control experiment **(B)**. Both functional MRS and control experiments begin with a BOLD-fMRI acquisition upon visual stimulation, to verify the presence of a BOLD response in the superior colliculus in the beginning of the session (Step 1). Then, spectra are acquired in the superior colliculus during ~1h10min, concatenating 5 repetitions of the ensemble Block Paradigm - Recovery (A) or No Stim 1 - No Stim 2 (B) (Step 2). Another BOLD-fMRI map is acquired to verify the presence of a BOLD response in the superior colliculus in the end of the MRS session (Step 3) for both (A) and (B). In the functional MRS experiment (A) only, functional MRS of water is then acquired in the SC.

**Experimental parameters**

**Steps 1 and 3:** BOLD fMRI experiments were acquired using a gradient echo EPI pulse sequence (TR/TE = 1000/12 ms), with an in-plane resolution of 0.145 × 0.145 mm$^2$ (axial, matrix size 110 x 96, FOV=16 x 13.9 mm$^2$) and a slice thickness of 0.5 mm. The slice package (8 interleaved slices) was positioned to cover the superior colliculus and the visual cortex. The anatomical image was acquired using a T2-weighted RARE sequence (TE/TR = 40/2000 ms, resolution: 0.075 × 0.080 × 0.400 mm$^3$, 15 interleaved slices).

**Step 2:** fMRS experiments were carried out using the adapted LASER sequence (described above) on a volume of 2.3 × 1.7 × 1.7 mm$^3$ voxel placed around the superior colliculus. Spectroscopic signals were collected over a bandwidth of 4000 Hz (10 ppm) with 2048 complex data points sampled (TE = 28 ms and TR = 1500 ms, total signal acquisition time = 512 ms). Hyperbolic secant (HS4) (Tannüs and Garwood, 1996) refocusing pulses (pulse duration = 2 ms, bandwidth = 10 kHz) were designed using Topspin 3.1 (Bruker). Water suppression was achieved using a CHESS (CHEmical Shift Selective saturation, (Frahm *et al.*, 1989)) module consisting of three CHESS pulses with a bandwidth of 250 Hz, chosen for having a shorter duration than the very efficient VAPOR (variable power RF pulses with optimized relaxation delays, (Tkáč *et al.*, 1999)). The water residual amplitude was usually lower than twice the N-acetylaspartate (NAA) amplitude. The water suppression pulses were interleaved with 4 blocks of outer volume suppression sinc10 pulses, giving a total duration for the preparation modules of 227 ms. For each BP-R paradigm (or No Stim 1 – No Stim 2 paradigm), 544 repetitions were acquired over 13 min 36 sec (x5). The single voxel was shimmed between each block of 13 min 36 s.

**Step 4:** Water fMRS was acquired using the same parameters as given above, only removing the water suppression module leading to 162 ms of preparation modules (OVS).

**Data Analysis.**

All data in this study were analyzed using custom code written in MATLAB® (The MathWorks, Inc., Natick, Massachusetts, United States). Below we elaborate on every data analysis routine for the relevant steps.

**Steps 1 and 3: BOLD-fMRI analysis**

The BOLD-fMRI map acquired in the beginning of the session was always analyzed on-the-fly with the vendor software (Paravision 6.0.1), to check if activity was detected in the superior colliculus before starting the MRS acquisitions. A more rigorous analysis with a home-made MATLAB® routine calling SPM12 was performed afterwards. Raw data were converted to Nifti, corrected for outliers, slice timing (taking for reference a slice comprising the superior colliculus) and motion (translation and rotation). Images were then isotropically smoothed using a 3D-Gaussian-kernel with FWHM=0.145 mm. Maps were fitted with a GLM (high-pass filter cutoff, 90s; motion parameters used as nuisance regressors) and thresholded at a significance level of p=0.01 and minimum cluster size=4.

**Step 2: fMRS analysis**

***Preprocessing.*** The phase cycling imposed a minimum time resolution of 4 repetitions (6 s). Repetitions were individually phase- and frequency-corrected taking as a reference the average spectrum with a MATLAB® home-made routine. For the sake of SNR, we summed spectra over 8 consecutive repetitions (each 4 repetitions) providing a moving average with an effective time resolution of 12 s (apparent time resolution 6 s). The resulting spectra were then eddy-current corrected in MATLAB® using the water reference (Klose, 1990) and the water residual signal (whose amplitude was usually not greater than twice the NAA or total creatine (tCr) peak) was removed by singular value decomposition (Pijnappel *et al.*, 1992) to ensure a flat baseline. To correct for the potential linewidth and intensity slow drifts along time (1 h 08 min of MRS acquisitions), we first concatenated the 5 blocks, normalized by their own intensity mean, to correct for potential gain variations between blocks. The data were then detrended by applying to successive spectra a linear linewidth correction, from the calculated averaged linewidth of the 3 singlets (NAA + tCr + choline compounds (tCho)), on individual blocks first (13 min 36 s) and then on the full time course (1 h 08 min). The average linewidths were computed by measuring the FWHM directly on the spectra, whose FIDs were zero filled prior to FT to improve the precision.

***Time course construction.*** To build the time course of interest (the block paradigm followed by the recovery period, 13 min 36 s), we summed the spectra coming from the corresponding time points from the 5 blocks (applying phase and frequency correction), giving a final time course of 135 points with an apparent time resolution of 6 s, and an effective time resolution of 12 s. Each spectrum representative of a time-point was calculated by adding the 8x5=40 individual repetitions, which allowed a robust metabolite quantification for individual animals.

***Correction for BOLD-induced line narrowing.*** Positive BOLD responses in active areas can induce an increase in T2/T2* and therefore a linewidth narrowing, which in turn can bias the signal quantification.

Under the hypothesis that BOLD effects would have a similar temporal dynamic on water and metabolite signals, we used the water fMRS measurements as a reference for BOLD-induced time courses, taking advantage of its high SNR. The linewidth and amplitude of the water signals were computed for the water fMRS experiments to spot the empirical timeline of the BOLD events (underlined in light grey on **Figure 2B/2C**). As an example, we corrected for the T2/T2* effect on water by applying a line-broadening of 0.4 Hz during the identified BOLD events. Temporal fluctuations of water linewidth due to BOLD activation were thus effectively removed (**Figure 2C**, right panel).

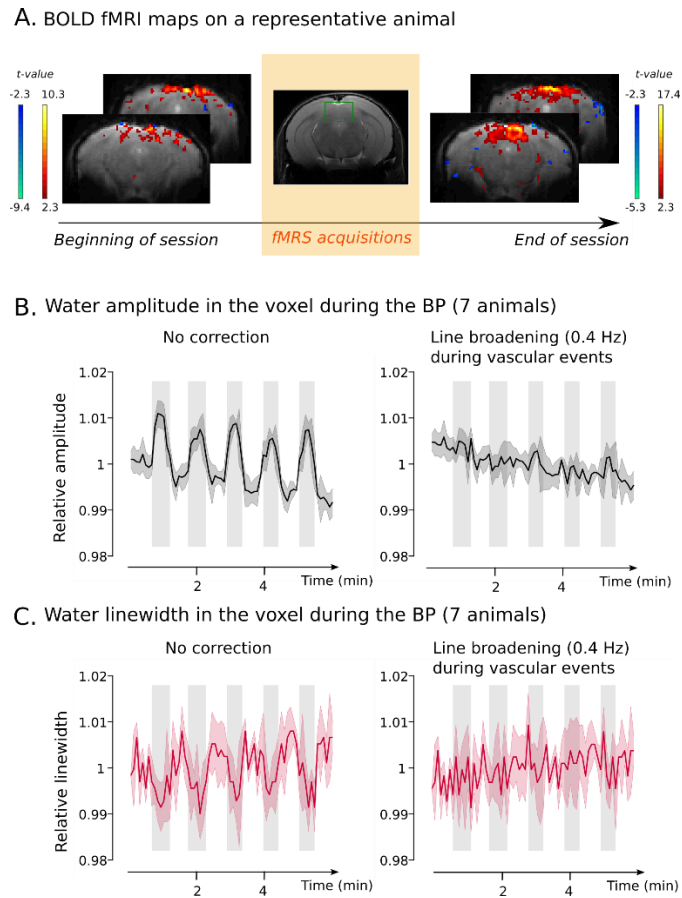

**Figure 2.** BOLD responses upon visual stimulation delivered in a block paradigm (BP). **(A)** Animal-averaged BOLD fMRI maps acquired in the beginning and in the end of experiments, with a significance threshold of p=0.01. **(B)** Animal-averaged time course of the water linewidth acquired in the voxel around superior colliculus during the block paradigm (Step 4) without any correction, used to define the BOLD events, enhanced in light grey (upper left). A line

broadening of 0.4 Hz is applied during the defined BOLD events to exemplify the effect of such a correction (upper right). Animal-averaged time course of the water amplitude acquired in the voxel around superior colliculus during the block paradigm (Step 4) without any correction (lower left) or with a line broadening of 0.4 Hz applied during the predefined BOLD events (lower right). The shadows represent the standard deviations.

The same approach was applied to the metabolites time courses (**Figure 3**). We computed the average linewidth and amplitude along the BP-R paradigm for the singlets of the upfield spectrum (NAA + tCr + tCho, **Figure 3B/3C**). We applied a line broadening of 0.5 Hz during the predefined BOLD events. Quantitatively, this correction equals at best the average linewidth during the BP and the average linewidth during the recovery period (**Figure 3C**). However, to account for cross-subject differences, we chose corrections individually for each animal (**Figure 3A**) by computing the difference between the averaged spectra of the BP and the Recovery periods, and applying a line broadening during the vascular events (between 0.5 and 0.7 Hz) that minimized the residuals of the 3 main singlets (**Supplementary Figure 1A**). Three (out of the 10) animals scanned for the functional experiment were detected as outliers: for these animals the nulling of the singlets' residuals required a line broadening > 1 Hz, corresponding to a linewidth drift independent of a BOLD effect. In the control group, two outliers were detected (**Supplementary Figure 1B**). To make both the functional and the control experiment comparable in number, we removed these outliers and one more control mouse which had the poorest spectral quality, leaving N = 7 mice in each group.

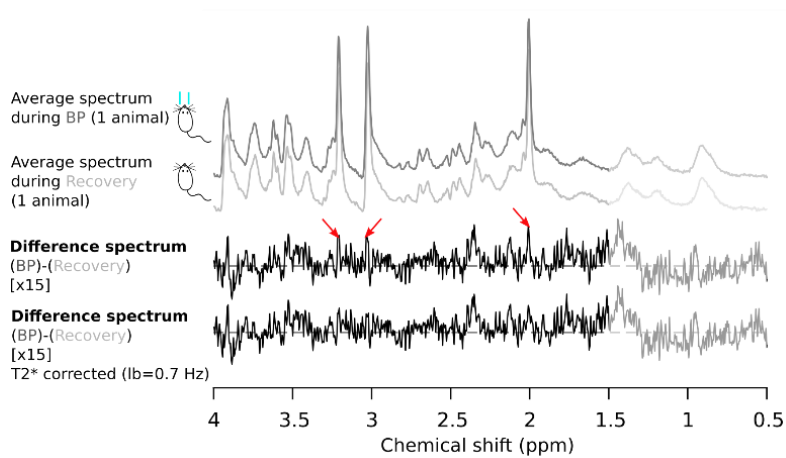

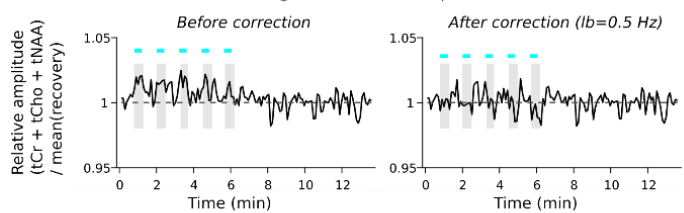

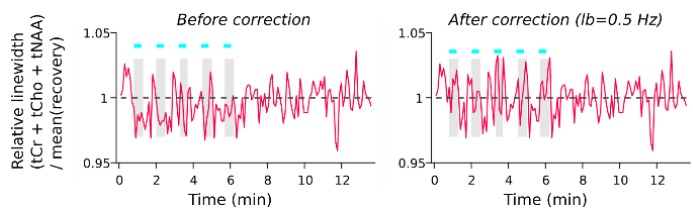

**Figure 3.** Correcting BOLD effects on metabolite signals. **(A)** Example of T2* correction on a representative animal. A difference spectrum is generated from the difference between the average spectrum during the Block Paradigm (BP) (dark gray) and the Recovery period (light gray). The T2* correction was applied during the predefined BOLD events. The line broadening was chosen to visually minimize the residuals of NAA, tCr and tCho. The region below 1.5 ppm is very variable, and should not be taken into account for interpretation. **(B)** Average amplitude of the 3 singlets (NAA+tCr+tCho) computed from the time course of the all-animal-averaged spectra. The correction consists in a line broadening of 0.5 Hz applied during the predefined BOLD events (light gray). The light blue horizontal bars represent the periods of visual stimulation. **(C)** Average linewidth of the 3 singlets (NAA+tCr+tCho) computed from the time course of the all-animal-averaged spectra. The correction consists in a line broadening of 0.5 Hz applied during the predefined BOLD events (light gray). The correction was chosen to obtain the same average linewidth during the block paradigm and the recovery period. The light blue horizontal bars represent the periods of visual stimulation.

***Time course quantification.*** The BOLD-corrected fMRS spectra were quantified using LCModel (Provencher, 1993). A home-made routine based on the density matrix formalism (Mulkern and Bowers, 1994) generated a basis set of 21 metabolites, using the chemical shift database proposed by (Govindaraju, Young and Maudsley, 2000). An empirical macromolecule baseline was included in the basis set for a more robust spectral quantification (Giapitzakis *et al.*, 2019). Macromolecule spectra were acquired during previous experiments in the same voxel by implementing a double inversion recovery module prior to the LASER localization block (first inversion time = 2200 ms, second inversion time = 700 ms, chosen for the resulting good apparent metabolic nulling, close to (Lopez-kolkovsky, Sebastien and Boumezbeur, 2016)). All other parameters were identical to the MRS acquisitions (Step 2) except the TR that had to be prolonged to 4000 ms to incorporate the double inversion module.

Spectra were fitted between 0.4 and 4.2 ppm. To obtain the normalized time course for every metabolite, the absolute concentration (as fitted by LCModel) was normalized by the corresponding average concentration during the recovery period (or during "No Stim 2" period for the control experiment) for every time point.

***Difference spectra.*** This analysis aimed at comparing the so-called "active period" (6 min 48 s of the BP) and the recovery period. We averaged spectra (with phase and frequency correction) from all time-points from the BOLD-corrected BP and subtracted the corresponding average spectrum from the recovery period. The same analysis was performed for the control experiment between the "No Stim 1" and "No Stim 2" periods.

# Results

Robust activation of the superior colliculus during the entire experimental duration

BOLD-fMRI maps acquired prior to the fMRS experiments and after their completion are shown in **Figure 2A**. Robust BOLD responses were observed in both instances, with similar spatial patterns. The contrast was occasionally somewhat weaker in the beginning of the experiment, likely because the isoflurane induction had not fully washed out yet. However, positive activation was always robustly observed in the superior colliculus, and in other visual areas such as the primary visual cortex and the lateral geniculate nucleus.

Moreover, the clear BOLD responses observed in the time courses of the functional MRS acquisitions on water (**Figure 2B**) suggest the volume of interest was well positioned and with little chemical shift displacement.

fMRS experiments - spectral quality

The SNR of a spectrum at one time point (representative spectrum, **Figure 4**) were 27±3 for the NAA resonance at 2.0 ppm (corresponding average linewidth = 13.2±1.8 Hz), 29±3 for the tCr resonance at 3 ppm (corresponding average linewidth = 12.9±1.1 Hz) and 28±3 for the tCho resonance at 3.2 ppm (corresponding average linewidth = 13.0±1.5 Hz). The chemical shift displacement between NAA (at 2 ppm) and tCr (at 3 ppm) equals 4% (that is to say, 0.09 mm in the axial plan and 0.07 mm in the coronal and sagittal plans). The individual CRLB (Cramer-Rao Lower Bounds) were smaller than 5% on individual time points for NAA, tCr, tCho, myo-inositol (Ins), Glu, taurine (Tau) and macromolecules. The baseline fitted by LCModel was generally very flat, suggesting a good quality of localization, water suppression and macromolecular fit.

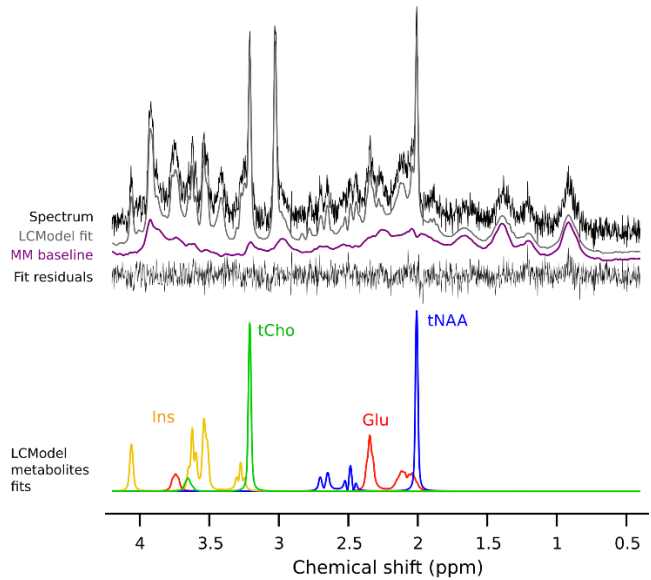

**Figure 4.** Representative spectrum of one time point (Recovery period) in one animal and the corresponding LCModel fit. The basis used for the linear combination incorporated 21 metabolites. The corresponding fit is shown for NAA, tCho (GPC + PCho + Cho), Glu, Ins.

<u>The difference spectrum reveals metabolic differences between the active period and the recovery period.</u>

To make the comparison possible between our results and the previous literature, we first considered the entire block paradigm as an "active" period, in contrast with the recovery period. The subtraction of the all-animal-averaged spectrum of the active period (corrected for vascular events as described in the Methods) and the all-animal-averaged spectrum of the recovery period is shown in **Figure 5B**, while the same analysis for the control experiment between No Stim 1 and No Stim 2 periods is shown in **Figure 5A**. The difference spectrum of the control experiment reveals non-vascular residuals for NAA, tCr and tCho, that can also be observed in the difference spectrum of the functional experiment. The principal difference between the two difference spectra is the robust residual at 2.3 ppm that we assign to glutamate. The region below 1.5 ppm was characterized by a large variability between animals (**Supplementary Figure 1**).

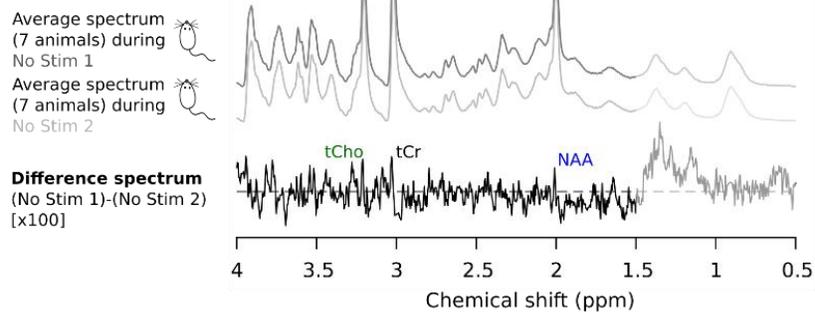

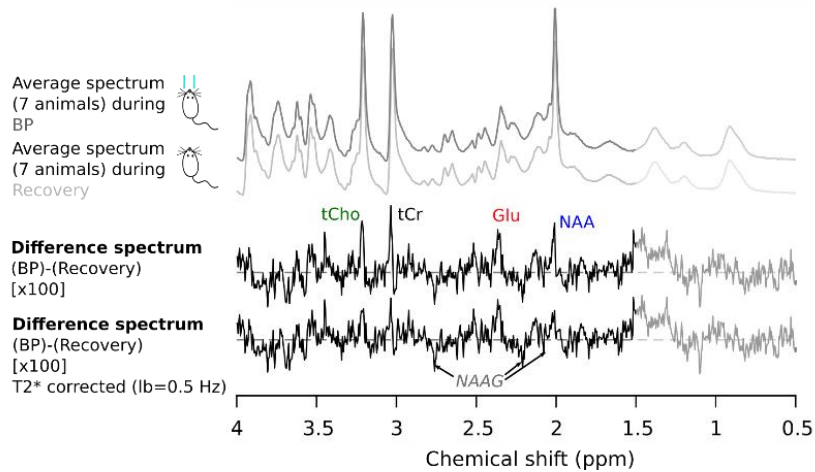

**Figure 5.** Spectral differences **(A)** Difference spectrum between the all-animal-averaged spectrum of the No Stim 1 period (dark grey) and the No Stim 2 period (light grey). The region below 1.5 ppm is very variable, and should not be taken into account for interpretation. **(B)** Difference spectrum between the all-animal-averaged spectrum of the Block Paradigm (dark grey) and the Recovery period (light grey). The region below 1.5 ppm is very variable, and should not be taken into account for interpretation

To compare the active period with the recovery period, we also computed the concentration distributions of metabolites for the BP and the recovery period, each value corresponding to 1 time point in 1 animal. Histograms for Glu, Ins, tCho and Glu are shown in **Figure 6B** (Tau and tCr in **Supplementary Figure 2**). Similarly, we compared No Stim 1 and No Stim 2 (**Figure 6C**). A consistent increase of glutamate during the block paradigm can be seen via a clear shift towards higher concentrations in the first part of the functional experiment, while there is no shift in the control distribution. Otherwise, a slight drift is bserved for NAA (and Tau) in both the functional and control experiments. tCr exhibits a stronger drift in the control experiment than in the functional experiment.

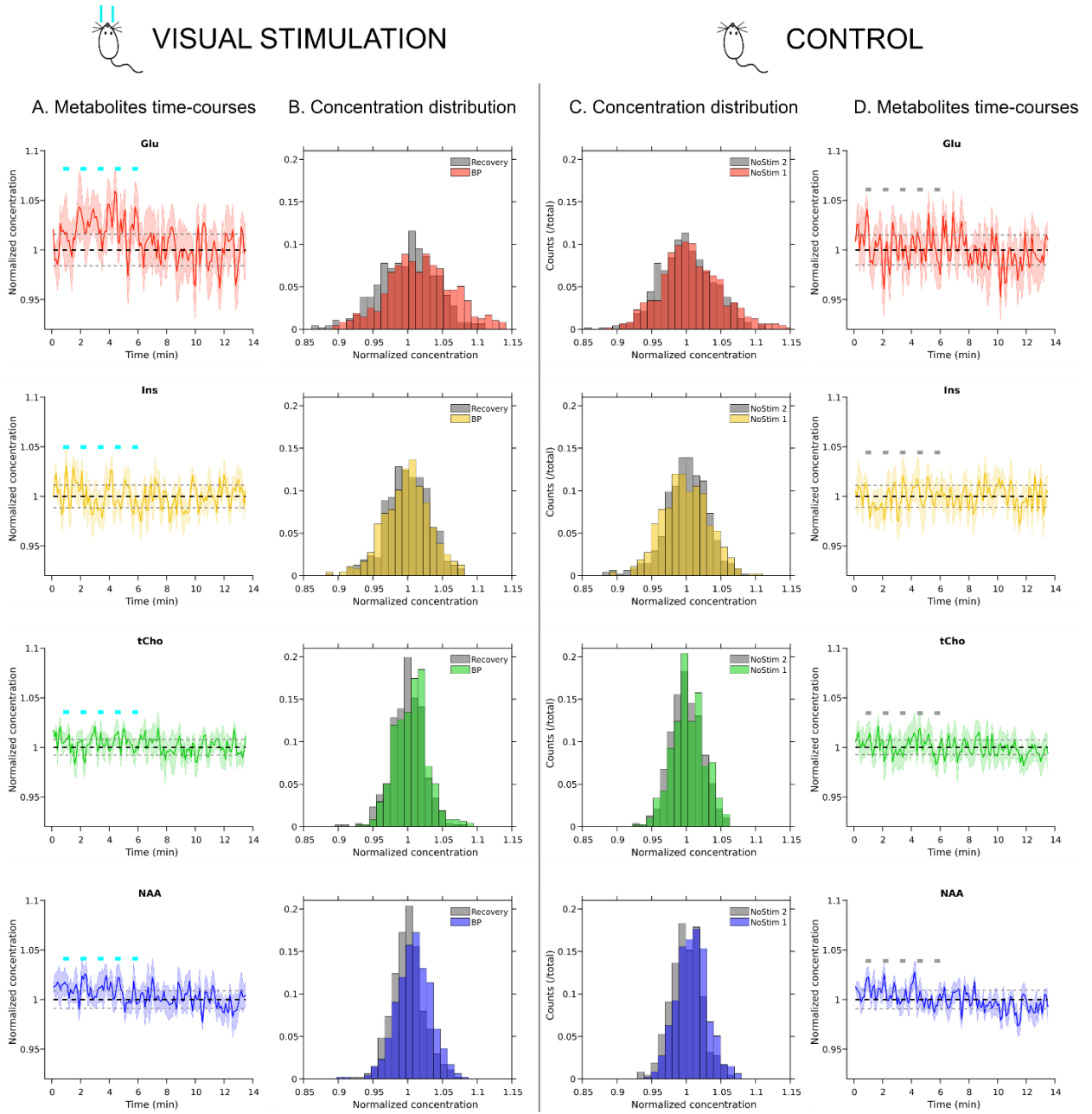

**Figure 6.** Time course analysis **(A)** Relative concentrations time-courses for Glu (red), Ins (yellow), tCho (green) and tNAA (blue) during the Block Paradigm (BP) and the Recovery period, normalized by their mean during the Recovery period. The dotted gray line represents the coefficient of variation of the data calculated from the standard deviation and the mean during the Recovery period. The light blue horizontal bars represent the periods of visual stimulation. **(B)** Relative concentrations distributions for Glu (red), Ins (yellow), tCho (green) and tNAA (blue) during the Block Paradigm (BP) and the Recovery period, normalized by their mean during the Recovery period.

Dynamic time courses of reliable metabolites (CRLB < 5%), time resolution 12s

The average time courses of NAA, tCr, tCho, Glu, Ins and Tau were computed from the individual time courses of the functional and control experiments (**Figure 6A, Supplementary Figure 2**). To account for the different noise levels of the different metabolites, we compare the metabolic variation with the coefficient of variation (CoV) defined during the recovery period (respectively the "No Stim 2" period) and represented it on the time courses by a dotted gray line. Glutamate is progressively upregulated during the visual stimulation. As for the inference of a more dynamic information, there does not seem to be an increase in glutamate locked to the short periods of stimuli (24 s, highlighted in light blue on **Figure 6A**), but the noise level prevents drawing more qualitative conclusions. The other metabolites present similar time courses between the functional and the control experiment, except for tCr, whose drift is stronger in the control experiment (**Supplementary Figure 2**).

Despite their much higher CRLB, time courses and distributions of N-acetylaspartateglutamate (NAAG) (mean CRLB = 10), PCr (mean CRLB = 7) and Cr (mean CRLB = 11) are displayed in **Supplementary Figure 3**, for their (qualitative) different patterns between the functional and the control experiment. Notably we notice a downregulation of NAAG during the block paradigm and an upregulation of creatine (Cr), and slight downregulation of phosphocreatine (PCr).

Generalized Linear Model (GLM) regression on time courses.

To get a more quantitative dynamic analysis, that the good time resolution theoretically allows, we ran a simple GLM regression on the average time courses of NAA, tCr, tCho, Glu, Ins and Tau. The GLM incorporated the block paradigm, the active/recovery design to assess for slow metabolic response, and a linear drift (**Figure 7C**). After a multiple comparison correction, NAA varied significantly with the block paradigm during the functional experiment, while Glutamate varied significantly only with the active/recovery design. During the control experiment, NAA, tCr, tCho exhibited a significant drift **(Table 1)**.

With a finite impulse response approach (13 blocks of 1min), we also estimated the "metabolic response function" (MRF) of the different metabolites, as it can be seen on **Figure 7A**. Glutamate MRF ramps up until the second period of stimulation (~3min) and decreases in the end of the BP. In comparison, the other metabolites do not exhibit MRF with a specific pattern.

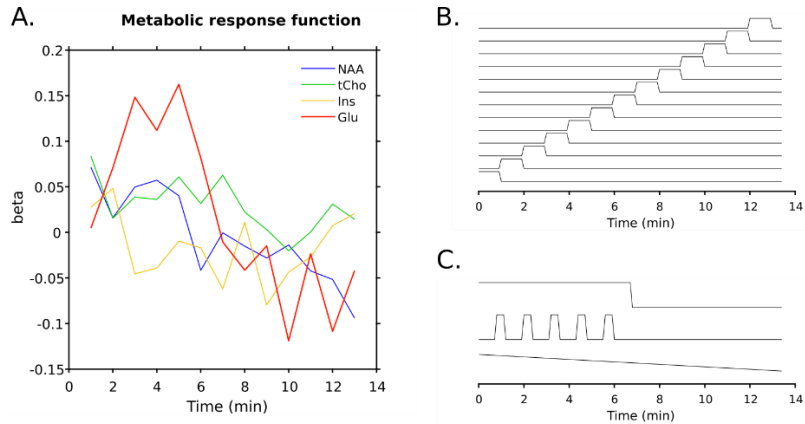

**Figure 7**. General Linear Model **(A)** Metabolic response function estimated for Glu, NAA, tCho and Ins with the Finite Impulse Response approach. **(B)** The regressor matrix for (A) contains 13 regressors. Each regressor consists in a step of ones of 1 min. **(C)** The regressor matrix corresponding to the GLM analysis whose results are shown in Table 1 contains 3 regressors: a linear drift (lower row), the block paradigm design (middle row) and active/recovery design (upper row).

## Discussion

In this work, we developed functional magnetic resonance spectroscopy experiments in mice. The mouse model is important for many disease models and is amenable to genetic manipulation; thereby, future studies using mouse fMRS could contribute to understanding of fMRS signals observed in humans and to interpret the ensuing metabolic variations. For example, several studies interpret variations of Glu, but more particularly GABA, as reporters of excitation or inhibition mechanisms, respectively (Cleve, Gussew and Reichenbach, 2015; Mekle *et al.*, 2017; Stanley and Raz, 2018; Bezalel, Paz and Tal, 2019; Boillat *et al.*, 2019), potentially making fMRS attractive for psychiatric diseases, where an altered excitation/inhibition balance is often pointed out (Marín, 2012). The availability of mouse models and their amenability towards concomitant or post-hoc invasive procedures, as well as the increasing availability of preclinical high field MRI scanners with high-performing probes can thus bridge the gap, and assist in explaining the nature of observed modulations. Optogenetics lines expressing channel rhodopsin specifically in glutamatergic or GABAergic neurons (Zhao *et al.*, 2011) could help disentangling the role of excitatory and inhibitory neurons in signal change (n.b. a recent study combining optogenetic and fMRS in rats (Just and Faber, 2019)). Moreover the well-established pharmacology available in rodents, and the chemical genetic lines (such as DREADDs, (Roth, 2016)) are preclinical tools allowing, among others, manipulations of diverse cell metabolic pathways or of synaptic release, and therefore have the potential to unravel the nature of the fMRS modulation. The increasing number of mouse models of psychiatric and neurodegenerative diseases reinforces the interest for translational approaches with functional magnetic resonance spectroscopy.

The robust increase of glutamate measured during the block paradigm is in good agreement with previous fMRS literature, in humans and in rodents. As the main excitatory neurotransmitter and abundant amino acid in the adult mammalian brain, glutamate is involved in numerous metabolic pathways (Danbolt, 2001; Magistretti and Allaman, 2015). It is mostly present in the cytoplasm, but is also densely located in synaptic vesicles, in mitochondria and in the synaptic cleft. Commonly associated to neurons in the CNS, glutamate also enters briefly in astrocytes through the glutamate-glutamine cycle. Therefore, we cannot assert the origin of the observed glutamate upregulation in our experiment. However, the GLM analysis (Table 1 and **Figure 7**) suggests that the modulation is not stimulation-locked, as would neurotransmission events with fast dynamics be, but rather reflects a relatively slow dynamic, possibly corresponding to a metabolic adaptation. Since the different glutamate pools might have different relaxation and diffusion properties, the upregulation can match an increased glutamate production, or a different compartmentalization.

The more subtle decrease of NAAG becomes quite consistent when adding the observations from the difference spectrum (**Figure 5B**) and from the time courses and distributions. NAAG is a neuropeptide particularly concentrated in the inhibitory layers of the SC (Tsai *et al.*, 1993), that may serve as an antagonist and a precursor of glutamate (Robinson *et al.*, 1987). Therefore, a parallel variation of NAAG and glutamate in this region is perhaps not very surprising.

Our relatively good spectral resolution disentangles PCr from Cr (with the 4ppm resonance). Some functional $^{31}$P MRS studies rather report an increase in the PCr signal during visual stimulation in the human brain. In a recent study (Hendriks *et al.*, 2019), the authors report an increase in PCr amplitude, but a decrease in linewidth during periods of stimulation, and therefore describe it as a BOLD effect. In (van de Bank *et al.*, 2018), the authors report an increase in the PCr amplitude, but no linewidth modulation during stimulation, claiming that PCr is upregulated during stimulation independently of a BOLD effect. However, PCr is a source of phosphate for ATP production through creatine kinase, and interestingly we observe a downregulation of PCr (and a consistent upregulation of Cr) more particularly in the beginning of the block paradigm. The phosphate of the PCr could be used to regenerate ATP, as a primary metabolic response. This result is consistent with a recent study achieved in the sedated tree shrew at 14.1 T (Sonnay *et al.*, 2018).

In comparison to several fMRS studies that report an increase of lactate during the active period (Mangia, Tkáč, Logothetis, *et al.*, 2007; Gussew *et al.*, 2010; Lin *et al.*, 2012; Schaller *et al.*, 2013; Just *et al.*, 2013; Bednařík *et al.*, 2015, 2018; Just and Sonnay, 2017; Mekle *et al.*, 2017; Boillat *et al.*, 2019; Koush *et al.*, 2019), we do not report such an increase. Several reasons can explain this difference. Most of the studies reporting a lactate increase during an active period delivered a sustained stimulus (> 2 min) or a prolonged task. In our work, the visual stimulus was delivered through a block paradigm with repeated short periods of visual stimulation (24s), and not through a sustained or very long stimulus, where lactate might accumulate in the cell before being cleared out. Additionally, the superior colliculus is a subcortical region of the brain, and previous studies usually observed the response in the cortical area associated to the sensory stimulus. In a recent rat fMRS study, metabolic changes upon a trigeminal nerve stimulation were observed and while an increase of lactate concentration in the somatosensory cortex was reported, similar thalamic variations were not reported (Just and Sonnay, 2017). The lactate increase is sometimes associated with the activity of the astrocyte-neuron (particularly excitatory) lactate shuttle. Any variation in cell-type distribution (astrocytes, inhibitory and excitatory neurons) may affect the underlying regional metabolism. We also cannot rule out that the anesthesia here used (specifically, a combination of medetomidine and isoflurane), could underlie differences in metabolic regimes. A recent study using medetomidine in rats (unlike most other studies that used alpha-chloralose) does not report changes in lactate during sensory or optogenetic stimulation (Just and Faber, 2019).

Our work points out the importance of a control experiment characterizing the time course of metabolites during long periods, as signal drifts may "masquerade" as activation (**Figure 6, Supplementary Figures 2 and 3**). Despite the shim drift correction, there might also be a metabolic drift, since anesthetics are affecting metabolism and measured metabolites concentrations (Boretius *et al.*, 2013).

The LCModel quantifies metabolites using the NMR signature area of each metabolite, that accounts for the linewidth shape, and therefore should be insensitive to T2* variation. However, some simple simulations (applying a variable line broadening on one spectrum) show that the concentration of some metabolites given by LCModel is affected by a linewidth difference of 0.5 Hz (up to 1-2%), and even more intensely with noisy spectra. Therefore, the correction for T2* effect (**Figure 2 and 3**) associated to the neurovascular coupling is a crucial postprocessing step, but the method is empirical and subsequently far from ideal. The line broadening in this study was applied to correct for T2* effect on NAA, tCho and tCr together; however, the linewidth of these metabolites might change slightly differently depending on their compartmentation and their proximity to vessels. Other non-vascular mechanisms could also change their T2 dynamically (relative change in compartmentation for example), suggesting that more robust T2* correction methods should be developed.

Finally, limitations associated to single voxel MRS and sedated animals also affect this study. The superior colliculus has a very heterogeneous structure, composed of 7 layers, with different types of neurons (May, 2006). The visual input arrives in the superficial layers of the SC (in the 3$^{rd}$ layer), and provokes an excitation followed by an inhibition from neurons in the 2$^{nd}$ layer. The volume of interest is placed around the complete structure (i.e. the 7 layers), and our measurements necessarily suffer from partial volume effect and heterogenous neuronal response. The use of sedation is also a confounding factor that might modulate metabolism and neurotransmission. Medetomidine is an α2-adrenergic agonist, however the density of α2-adrenergic receptors is quite low in the SC (Wang *et al.*, 1996), so the effect should be mostly indirect. Isoflurane binds to many receptors and its action mechanism is not clearly established, however we use it at very low dose (0.3%) to stabilize the medetomidine sedation, moderating its impact.

# Conclusion

Functional magnetic resonance spectroscopy was successfully developed and applied in the sedated mouse. Consistent glutamate increases were measured in the superior colliculus upon visual stimulation, delivered with a block paradigm that can be equally used for BOLD-fMRI. The effective time resolution (12 s) and our paradigm design (discontinuous stimulation) enabled a dynamic analysis with a GLM commonly used in fMRI analysis. Under our experimental conditions, the glutamate modulation measured by MRS is not locked to the stimulus and is likely partially independent from the neurotransmission events. This study opens the door for the exploration and application of fMRS in the mouse, the richest mammalian animal model in terms of genetic lines and pharmacological tools.

# Acknowledgements

This study was funded in part by the European Research Council (ERC) (agreement No. 679058). The authors acknowledge the vivarium of the Champalimaud Centre for the Unknown, a facility of CONGENTO which is a research infrastructure co-financed by Lisboa Regional Operational Programme (Lisboa 2020), under the PORTUGAL 2020 Partnership Agreement through the European Regional Development Fund (ERDF) and Fundação para a Ciência e Tecnologia (Portugal), project LISBOA-01-0145-FEDER-022170.

**Figures captions**

**Figure 1**. Representative timelines of the functional MRS experiment **(A)** and the control experiment **(B)**. Both functional MRS and control experiments begin with a BOLD-fMRI acquisition upon visual stimulation, to verify the presence of a BOLD response in the superior colliculus in the beginning of the session (Step 1). Then, spectra are acquired in the superior colliculus during ~1h10min, concatenating 5 repetitions of the ensemble Block Paradigm - Recovery (A) or No Stim 1 - No Stim 2 (B) (Step 2). Another BOLD-fMRI map is acquired to verify the presence of a BOLD response in the superior colliculus in the end of the MRS session (Step 3) for both (A) and (B). In the functional MRS experiment (A) only, functional MRS of water is then acquired in the SC.

**Figure 2.** BOLD responses upon visual stimulation delivered in a block paradigm (BP). **(A)** Animal-averaged BOLD fMRI maps acquired in the beginning and in the end of experiments, with a significance threshold of $p=0.01$. **(B)** Animal-averaged time course of the water linewidth acquired in the voxel around superior colliculus during the block paradigm (Step 4) without any correction, used to define the BOLD events, enhanced in light grey (upper left). A line broadening of 0.4 Hz is applied during the defined BOLD events to exemplify the effect of such a correction (upper right). Animal-averaged time course of the water amplitude acquired in the voxel around superior colliculus during the block paradigm (Step 4) without any correction (lower left) or with a line broadening of 0.4 Hz applied during the predefined BOLD events (lower right). The shadows represent the standard deviations.

**Figure 3.** Correcting BOLD effects on metabolite signals. **(A)** Example of T2* correction on a representative animal. A difference spectrum is generated from the difference between the average spectrum during the Block Paradigm (BP) (dark gray) and the Recovery period (light gray). The T2* correction was applied

during the predefined BOLD events. The line broadening was chosen to visually minimize the residuals of NAA, tCr and tCho. The region below 1.5 ppm is very variable, and should not be taken into account for interpretation. **(B)** Average amplitude of the 3 singlets (NAA+tCr+tCho) computed from the time course of the all-animal-averaged spectra. The correction consists in a line broadening of 0.5 Hz applied during the predefined BOLD events (light gray). The light blue horizontal bars represent the periods of visual stimulation. **(C)** Average linewidth of the 3 singlets (NAA+tCr+tCho) computed from the time course of the all-animal-averaged spectra. The correction consists in a line broadening of 0.5 Hz applied during the predefined BOLD events (light gray). The correction was chosen to obtain the same average linewidth during the block paradigm and the recovery period. The light blue horizontal bars represent the periods of visual stimulation.

**Figure 4.** Representative spectrum of one time point (Recovery period) in one animal and the corresponding LCModel fit. The basis used for the linear combination incorporated 21 metabolites. The corresponding fit is shown for NAA, tCho (GPC + PCho + Cho), Glu, Ins.

**Figure 5.** Spectral differences **(A)** Difference spectrum between the all-animal-averaged spectrum of the No Stim 1 period (dark grey) and the No Stim 2 period (light grey). The region below 1.5 ppm is very variable, and should not be taken into account for interpretation. **(B)** Difference spectrum between the all-animal-averaged spectrum of the Block Paradigm (dark grey) and the Recovery period (light grey). The region below 1.5 ppm is very variable, and should not be taken into account for interpretation

**Figure 6.** Time course analysis **(A)** Relative concentrations time-courses for Glu (red), Ins (yellow), tCho (green) and tNAA (blue) during the Block Paradigm (BP) and the Recovery period, normalized by their mean during the Recovery period. The dotted gray line represents the coefficient of variation of the data calculated from the standard deviation and the mean during the Recovery period. The light blue horizontal bars represent the periods of visual stimulation. **(B)** Relative concentrations distributions for Glu (red), Ins (yellow), tCho (green) and tNAA (blue) during the Block Paradigm (BP) and the Recovery period, normalized by their mean during the Recovery period.

**Figure 7**. General Linear Model **(A)** Metabolic response function estimated for Glu, NAA, tCho and Ins with the Finite Impulse Response approach. **(B)** The regressor matrix for (A) contains 13 regressors. Each regressor consists in a step of ones of 1 min. **(C)** The regressor matrix corresponding to the GLM analysis whose results are shown in Table 1 contains 3 regressors: a linear drift (lower row), the block paradigm design (middle row) and active/recovery design (upper row).

**Table 1**: p-values for the GLM analysis achieved on metabolites time courses for both the functional and the control experiment, whose regressors are described in Figure 7C. p-values < 0.05 are highlighted in **bold**. When applying a Bonferroni correction for multiple comparison (6 metabolites), the p-values becoming > 0.05 are highlighted in ***bold + italic***.

|      | FUNCTIONAL | | | CONTROL | | |
| --- | --- | --- | --- | --- | --- | --- |
|      | Drift | Block Paradigm | Active/Recovery | Drift | Block Paradigm | Active/Recovery |
| NAA  | 0.019 | **0.003** | 0.081 | **1.07E-05** | 0.131 | 0.24 |
| tCr  | 0.116 | ***0.015*** | 0.567 | **2.54E-04** | 0.817 | 0.718 |
| tCho | 0.756 | 0.133 | 0.552 | **0.00498** | 0.14 | 0.548 |
| Glu  | 0.812 | 0.512 | **0.0041** | ***0.0451*** | 0.053 | 0.834 |
| Ins  | 0.467 | 0.222 | 0.733 | 0.35 | 0.989 | 0.539 |
| Tau  | 0.122 | 0.629 | 0.67 | 0.206 | ***0.025*** | 0.371 |